# Single-image example-based superresolution of hyperpolarized $^{13}$C images

Kofi Deh[1], Elizabeth Coffee[2], Guannan Zhang[1], Nathaniel Kim[1], Vesselin Miloushev[1], Kayvan R. Keshari[1,3,4,5*]

[1]Department of Radiology, Memorial Sloan Kettering Cancer Center, New York City, NY, USA

[2]Department of Neurology, Memorial Sloan Kettering Cancer Center, New York City, NY, USA

[3]Department of Physiology & Biophysics, Weill Cornell Graduate School, New York City, NY, USA

[4]Department of Biochemistry & Structural Biology, Weill Cornell Graduate School, NYC, NY, USA

[5]Molecular Pharmacology Program, Memorial Sloan Kettering Cancer Center, NYC, NY, USA

## Abstract

The transient signal available for hyperpolarized (HP) $^{13}$C imaging limits the image spatial resolution. This paper demonstrates a method of increasing spatial resolution by post-processing the HP $^{13}$C metabolic map using a corresponding anatomic image as a high-resolution example. The proposed method is tested with a Shepp-Logan digital phantom and for an MR image of a tumor in a mouse brain acquired at 3 Tesla using $^{13}$C Chemical Shift Imaging. The results are compared to the results of standard image rescaling methods using peak-signal-to-noise ratio and structural similarity index (SSIM) image quality metrics. In vivo, the method reduced blur resulting in a better match between the metabolite biodistribution and the anatomy. It is concluded that the image quality of hyperpolarized $^{13}$C MRI *in vivo* images can be improved with single image superresolution using the anatomic image as an example.

**Introduction**

Dissolution dynamic nuclear polarization (dDNP) greatly enhances the sensitivity of magnetic resonance imaging (MRI) making it a powerful tool for metabolic imaging [1]. Despite this enhancement, the signal-to-noise ratio (SNR) is still much less than is obtained for proton imaging because the concentration of the injected $^{13}$C pyruvate is in the millimolar range as compared to approximately 110 molar for protons. Therefore, $^{13}$C metabolic images are acquired at a low spatial resolution to obtain sufficient SNR, rescaled and fused with a proton anatomic image to facilitate visualization of a metabolite's biodistribution.

Multiple methods have been implemented for image rescaling including nearest neighbor interpolation which clones each image pixel to maintain image edges, and image domain sinc-shaped interpolation, such as bicubic and Lanczos-3 methods, which maintains a smooth intensity continuation. To remedy the image blurring introduced by these methods, the use of prior knowledge of the edges in a high-resolution anatomic (usually $T_1$-weighted or $T_2$-weighted) image for enhancing edges in the low-resolution image has been proposed [2, 3]. This approach has been adapted for proton magnetic resonance spectroscopic imaging (MRSI) and subsequently HP $^{13}$C imaging; however, as noted by authors, the quality of the results is subject to the similarity in contrast between the anatomic and metabolic images, the accuracy of segmentation of the proton anatomic image, and the complexity of the anatomy. [4-7]. In this paper, we propose that the reference anatomic image should be acquired with the same imaging sequence that was used for acquiring the metabolic image. This results in increased similarity between the anatomic and metabolic image, allowing the use of an example-based super-resolution method, a method that avoids segmentation, to upscale the metabolic image.

**Methods**

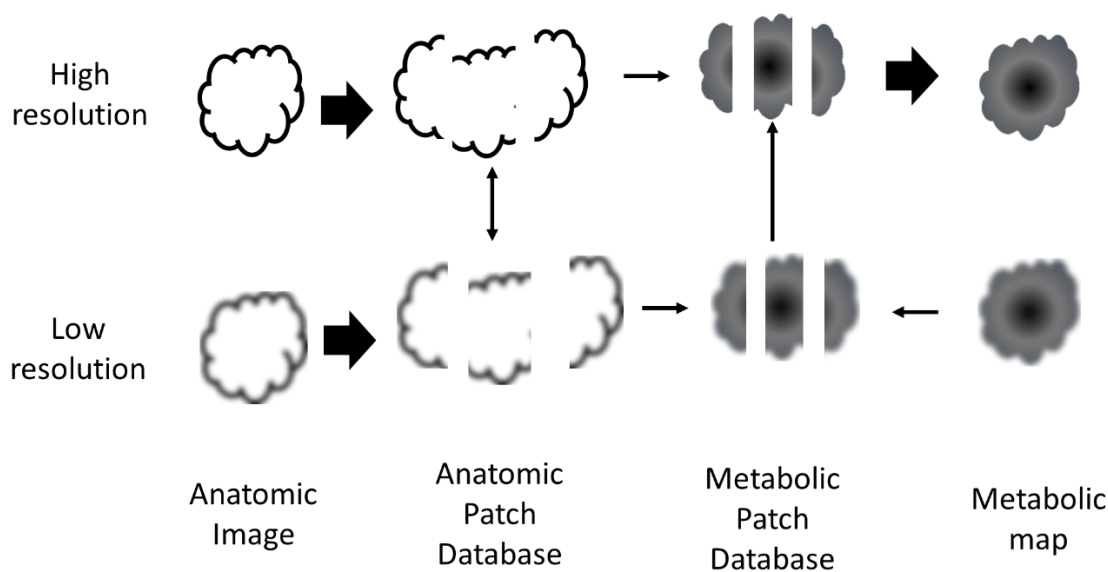

Figure 1 Illustration of superresolution of metabolic map using anatomic image as an example. Corresponding patch databases are built from the low- and high-resolution images. Patches from the high-resolution anatomic database are that best match patches from a corresponding low-resolution metabolic map are combined to form are high-resolution metabolic map.

The example-based super-resolution algorithm is described in **[8]** and our implementation is described below. The method is based on the principle of cross-scale invariance, which states that certain image features are preserved regardless of spatial resolution [8]. In our implementation, we generate a training data set by extracting overlapping patches from a high-resolution anatomic image. The overlaps ensure that each patch for reconstructing a new high-resolution metabolic image will come from a location in the low-resolution image that is compatible with its neighboring patches. The proton anatomic image is ideally acquired with the same sequence acquisition parameters used for $^{13}$C imaging. However, the high-resolution anatomic image that is routinely acquired to facilitate visualization of metabolite biodistribution can be downscaled to the size of the metabolic image for example-based super-resolution, as is shown in our demonstration. A low-resolution image, $Y$, and its high-resolution version, $X$, acquired by the same scanner can be modeled as,

$$Y = AX + n \qquad (1)$$

where $n$ is the image acquisition noise. $A$ can be an identity in which case solving for $X$ becomes a denoising problem, but in general it is a linear degradation operation that includes downsampling and blurring operations. $X$ can be approximately recovered from $Y$, using a linear interpolation filter, $\mathcal{L}$, such as nearest neighbor or bicubic interpolation. This filter adequately transfers low spatial frequency features from $Y$ to $X$, leaving the problem of how to transfer high spatial frequencies, i.e., edge information. Assume $X_t$ and $Y_t$ are corresponding high- and low-resolution anatomic images of a subject acquired in an HP $^{13}$C experiment. A high-pass filter, $\mathfrak{h}(\cdot)$, can be used to obtain a map of the missing edges, $Y_t^\dagger$, in the linearly interpolated low-resolution image by subtracting this image from the edge map of the high-resolution image:

$$Y_t^\dagger = \mathfrak{h}\big(\mathcal{L}(Y_t)\big) - \mathfrak{h}(X_t) \qquad (2)$$

After extracting a set of overlapping patches, $P = \{y_t^\dagger \in Y_t^\dagger, x_t^\dagger \in X_t^\dagger\}$, from the low- and high-resolution edge maps, a unique identifier, $k_i$, can be created for each high-resolution edge patch by concatenating its neighboring pixels, $\mathcal{N}\{y_t^\dagger\}$, with pixels from a low-resolution edge patch

$$k_i = \mathcal{N}(x_t^\dagger) \cup y_t^\dagger \qquad (3)$$

With these identifiers, a regularization scheme, $R(k)$, can be defined to enforce the spatial consistency when recovering $x^\dagger$:

$$R(k) = \sum_{k_j \in P} \|k_j - k_i\|_2^2 \qquad (4)$$

To find the optimal high-resolution edges for a newly acquired low-resolution image, an edge map that satisfies

$$\widehat{x^\dagger} = \underset{x^\dagger}{\mathrm{argmin}} \|x^\dagger - y^\dagger\|_2^2 + \lambda R(k) \qquad (5)$$

can be constructed. Here, $\lambda$ is a user-defined constant that controls the trade-off between fidelity to the edges in the acquired low-quality image and incorporation of prior knowledge from the edges in the high-quality image. After restoring each patch in the edge map, the new edge map is used to replace the original edge map in the acquired low-resolution image:

$$\hat{x} = \widehat{x^\dagger} + (\mathcal{L}(y)) - \mathfrak{h}(\mathcal{L}(y)) \qquad (6)$$

A MATLAB implementation of this method can be found at github.com/skmd3538/HP13CSR. We compare the results from applying the proposed technique to other image interpolation methods described previously, bicubic (BL), nearest neighbor (NN) and Lanczos-3 (LZ3) interpolation, using the peak-signal-to-noise ratio (PSNR) and the structural similarity index (SSIM) metrics. The PSNR is the mean squared error between the images being compared, scaled to compensate for any differences in intensity ranges. Since PSNR is a global assessment of image statistics, it does not accurately capture human perception of image fidelity which is affected by local features [9, 10]. This is achieved by the structural similarity index (SSIM), which compares differences in the brightness, contrast, and structure between the images. The two image comparison metrics are calculated in this study using *psnr()* and *SSIM()* functions in MATLAB (Mathworks).

**Digital Phantom Experiments**

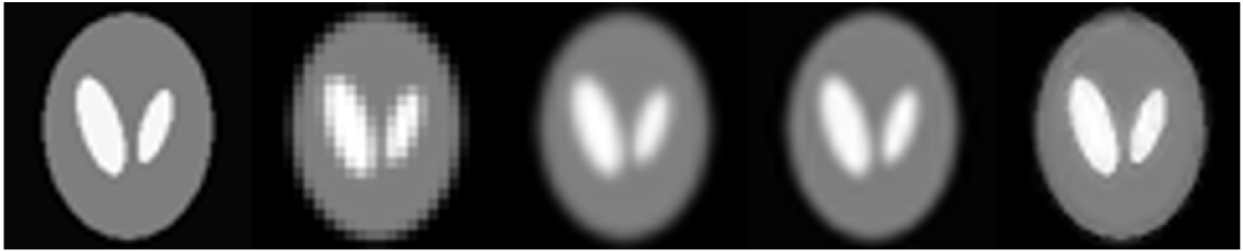

|              | Anatomic Image | NN    | BL    | LZ3   | IQT   |
|--------------|----------------|-------|-------|-------|-------|
| PSNR         |                | 0.090 | 0.092 | 0.093 | 0.095 |
| SSIM         |                | 7.089 | 7.028 | 7.063 | 7.394 |

Figure 2 Comparison of the performance of the proposed method (IQT) to standard interpolation methods for upscaling of a Shepp-Logan "metabolic" phantom. The "metabolic" phantom was created by downscaling a Shepp-Logan phantom by a factor of 2 and adding Gaussian blur with a standard deviation of 0.5. The results of upscaling this "metabolic phantom" by nearest neighbor (NN), bilinear (BL), Lanczos-3(LZ3) and IQT are shown with similarity of the result (PSNR & SSIM) to the original ("anatomic") phantom.

We tested the proposed method in a digital phantom. **Figure 2** shows an "anatomic" image, a 48 x 48 Shepp-Logan digital phantom created using MATLAB's phantom() command. A "metabolic" phantom is generated by bilinear interpolation of the "anatomic" phantom to a size of 24 x 24 followed by addition of Gaussian blur with a standard deviation of 0.5. Corresponding patch databases of the high-resolution "anatomic" and the low-resolution "metabolic" images were created as described in Figure 1, and the high-resolution "metabolic" image is reconstructed by selecting the set of patches from the high-resolution "anatomic" database using the patches in the low-resolution "metabolic" phantom as keys.

**Preclinical Experiments**

The proposed method was applied to the super-resolution of HP $^{13}$C images acquired with Chemical Shift Imaging (CSI) on a preclinical Bruker Biospec 3T MRI scanner (maximum gradient strength = 959 mT/m, maximum slew rate = 6393 T/m/s), using a dual-tuned transmit/receive $^1$H/$^{13}$C 42 mm diameter birdcage coil. Hyperpolarization of all chemical substrates was performed in a 5T SPINlab hyperpolarizer (GE Healthcare) for approximately 2 hours at 5T and 0.8K. *In vivo* CSI spectra of the mouse brain were acquired using the CSI sequence with a matrix size of 12 x 12 and FOV of 34 x 34 mm after injection of $^{13}$C-pyruvate into the tail-vein of the mouse. An axial T$_2$ weighted image of the mouse brain was acquired at a resolution of 0.13 mm

**Image Reconstruction**

The FIDs from the CSI data acquisition was Fourier-transformed, baseline and phase-corrected to obtain spatially localized maps of $^{13}$C spectra [11]. Intensity maps of metabolites were generated by integrating the magnitude of the spectra for the range corresponding to each metabolite's chemical shift plus or minus one. Custom MATLAB code was used for super-resolution of phantom and mice images. The IQT weights were generated sequentially in two steps each with a scale factor of 2 using a low-resolution patch of 7 x 7 pixels and a high-resolution patch size of 5 x 5 pixels. The patch sizes are selected based on results from previous authors showing these sizes resulted in a good trade-off between performance and computational time [12]. For comparison, the low-resolution image was also upsampled using nearest neighbor, Lanczos-3 and bicubic interpolation. For digital phantom experiments, the PSNR and SSIM between the upsampled images and the example anatomic high-resolution image were calculated.

**Results**

We applied the proposed method to digital and in vivo images. **Figure 2** shows the results of upsampling the "metabolic" using the standard image interpolation techniques and the proposed method, which we call image quality transfer (IQT). The nearest neighbor interpolation (NN) is the most accurate replication of the "metabolic phantom", while the bilinear (BL) and Lanczos-3 (LZ3) introduce smoothness to the interpolated image at the cost of increased blur. IQT is able to remove this blur resulting in the best match between the upscaled "metabolic" and "anatomic" image.

**Figure 3A** shows a $T_2$-weighted MR image of a mouse brain with a tumor, while **Figure 3B** (NN) is the nearest neighbor interpolation of a metabolic map of the brain after [1-$^{13}$C]-pyruvate injection. The metabolic map was acquired at a matrix size of 12 x 12, zerofilled to 24 x 24 before being upscaled to 96 x 96. IQT reduces blur making the visualization of the tumor easier.

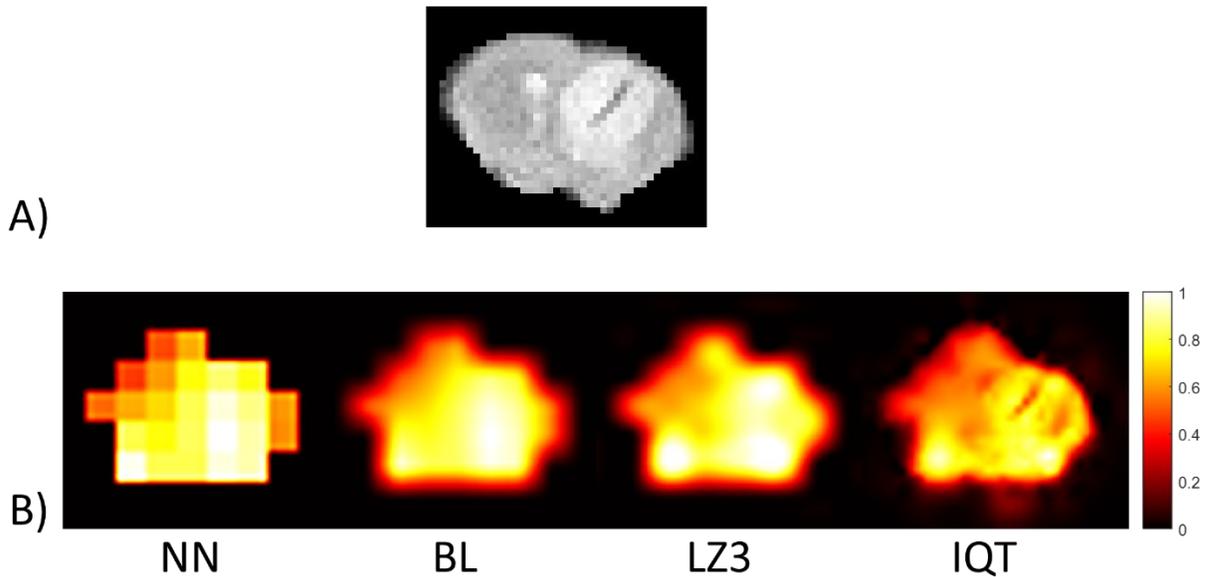

Figure 3 A) A T2-weighted MRI of a mouse brain with a tumor. The image matrix size was 128 x 128. The image is downscaled to a size of 12 x 12 for a low-resolution example and 96 x 96 for a high-resolution example. B) A 12 x 12 size 13C map is zerofilled to 24 x 24 and then upscaled to 96 x 96 using nearest (NN), bilinear (BL), Lanczos-3(LZ3) interpolation, and the proposed method (IQT).

**Discussion**

Hyperpolarized $^{13}$C imaging provides a practical way for implementing magnetic resonance imaging of metabolic flux in the clinic. However, improvements in SNR and spatial resolution of the acquired images are necessary to increase their utility. In this work, we demonstrate the use of example-based super-resolution for improving the spatial resolution of the metabolic map using a corresponding anatomic map as an example. Unlike other post-processing methods proposed for this task, our method does not require registration or segmentation of an anatomic image and can be used for any region of the anatomy. We showed that, compared to linear interpolation methods, the proposed method generates images of phantoms with higher PSNR and SSIM values than are obtained using linear interpolation methods. We also showed that, in vivo, it produces metabolic maps with reduced blur compared to when the maps are upscaled with commonly used image rescaling methods.

A limitation of this study is the lack of comparison to an in vivo ground truth. This is because it is virtually impossible to acquire an invivo high-resolution HP $^{13}$C image because of the limited duration of the signal. However, future work will investigate comparisons with ex vivo metabolic images.

In conclusion, we have demonstrated a method for improving the spatial resolution of metabolic maps using an anatomic image as an example.

**Acknowledgements:** This work was sponsored by the National Institutes of Health (NIH R01 CA195476, R01 CA237466, R01 CA252037, R01 CA248364 and S10 OD016422; NIH/NCI Cancer Center Support grant P30 CA008748); Center for Molecular Imaging and Nanotechnology at MSKCC; The Thompson Family Foundation; The Center for Experimental Therapeutics at MSKCC; Mr. William H. and Mrs. Alice Goodwin and the Commonwealth Foundation for Cancer Research; and the Peter Michael Foundation. Unrelated to this work, K.R.K. is co-founder of Atish Technologies and serves on the scientific advisory board of NVision Imaging Technologies and Imaginostics. He is a named inventor on patents related to imaging of cellular metabolism that are not related to this work.